\documentclass{sig-alternate-2013}

\usepackage{graphicx}
\usepackage{numprint}
\usepackage{subfig}
\usepackage{url}

\usepackage[font=bf,skip=\baselineskip]{caption}

\usepackage{microtype}

\usepackage{balance}

\usepackage[hyperfootnotes=false]{hyperref}

\usepackage[utf8x]{inputenc} 
\usepackage[X2,T1]{fontenc} 
\usepackage{layouts}
\usepackage{textcomp}
\usepackage{dblfloatfix}
\usepackage{placeins}
\usepackage[numbers,sort]{natbib}

\newcommand{\para}[1]{\smallskip\noindent\textbf{#1}}

\begin{document}

\newcommand{\helv}[1]{{\fontfamily{helv}\selectfont #1}}

\title{Sampling from Social Networks with Attributes}

\makeatletter
\def\@copyrightspace{\@float{copyrightbox}[b]
\begin{center}
\setlength{\unitlength}{1pc}
\begin{picture}(20,5.5) %
\put(0,-0.95){\crnotice{\@toappear}}
\end{picture}
\end{center}
\end@float}
\makeatother

\permission{\copyright 2017 International World Wide Web Conference Committee \\ (IW3C2), published under Creative Commons CC BY 4.0 License.}
\conferenceinfo{WWW 2017,}{April 3--7, 2017, Perth, Australia.}
\copyrightetc{ACM \the\acmcopyr}
\crdata{978-1-4503-4913-0/17/04. \\
	http://dx.doi.org/10.1145/3038912.3052665 \\
	\includegraphics[scale=0.8]{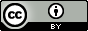}}
\numberofauthors{5}
\author{
\alignauthor
Claudia Wagner\thanks{Both authors contributed equally to this work.\vspace{-2em}}\\
      \affaddr{\scalebox{.99}[1.0]{GESIS \& U. of Koblenz-Landau}}\\
      \email{claudia.wagner@gesis.org}
\and
\alignauthor
Philipp Singer\footnotemark[1]\\
      \affaddr{\scalebox{.99}[1.0]{GESIS \& U. of Koblenz-Landau}}\\
      \email{philipp.singer@gesis.org}
\and
\alignauthor
Fariba Karimi\\
      \affaddr{\scalebox{.99}[1.0]{GESIS \& U. of Koblenz-Landau}}\\
      \email{fariba.karimi@gesis.org}
\and
\alignauthor
J\"{u}rgen Pfeffer\\
      \affaddr{Technical University of Munich }\\
      \email{juergen.pfeffer@tum.de}
\and
\alignauthor
Markus Strohmaier\\
      \affaddr{\scalebox{.99}[1.0]{GESIS \& U. of Koblenz-Landau}}\\
      \email{markus.strohmaier@gesis.org}
}

\maketitle

\begin{abstract}
Sampling from large networks represents a fundamental challenge for social network research. In this paper, we explore the sensitivity of different sampling techniques (node sampling, edge sampling, random walk sampling, and snowball sampling) on social networks with attributes. We consider the special case of networks (i) where we have one attribute with two values (e.g., male and female in the case of gender), (ii) where the size of the two groups is unequal (e.g., a male majority and a female minority), and (iii) where nodes with the same or different attribute value attract or repel each other (i.e., homophilic  or heterophilic behavior). We evaluate the different sampling techniques with respect to conserving the position of nodes and the visibility of groups in such networks. Experiments are conducted both on synthetic and empirical social networks. Our results provide evidence that different network sampling techniques are highly sensitive with regard to capturing the expected centrality of nodes, and that their accuracy depends on relative group size differences and on the level of homophily that can be observed in the network. We conclude that uninformed sampling from social networks with attributes thus can significantly impair the ability of researchers to draw valid conclusions about the centrality of nodes and the visibility or invisibility of groups in social networks.
\end{abstract}

\vspace{1mm}

\section{Introduction}

Sampling from large networks represents a fundamental problem for social network research. %
In order to draw valid conclusions from network samples, understanding how accurately samples reflect the position of nodes in the original network is essential.
Previous research has studied robustness of network samples from different angles, for example by examining the accuracy of network measures such as degree or betweenness centrality. A range of network properties has been found to be sensitive to the choice of sampling methods \cite{leskovec2006sampling,Galaskiewicz1991,Costenbader2003,Huisman2009,Borgatti2006,Kossinets2006,wang2012measurement,Lee2015}.

\para{Motivation and problem.} In this paper, we focus on the specific problem of sampling nodes and edges from a social network with attributes, i.e., a network where nodes are colored. For example, the color of nodes might be determined by gender, ethnicity, or age. We consider the special case of networks (i) where one binary attribute can be observed (e.g., a male and a female group of nodes), (ii) where the size of the two groups is unequal (e.g., a male majority and a female minority), and (iii) where nodes with the same or different attribute value attract or repel each other, i.e., homophilic \cite{shrum1988friendship} or heterophilic networks \cite{bearman2004chains}.
While the general impact of sampling on network characteristics has been studied thoroughly in the past \cite{Galaskiewicz1991,Costenbader2003,Huisman2009,Borgatti2006,Kossinets2006,wang2012measurement,Lee2015}, the role of attributes in combination with fundamental social mechanisms such as homophily \cite{mcpherson2001birds,Simsek2008}
has only received little attention so far \cite{Li2011}. In fact little is known about whether or how different sampling techniques are able to conserve the \emph{ranking of nodes} or the \emph{visibility of groups} from the original network.
Accurately capturing network characteristics of groups of nodes in sampled data, however, is crucial not only for researchers interested in directly studying these groups (e.g., gender or sociological studies), but also for researchers interested in analyzing the structure of the complete network since attributes of actors can impact the overall network structure \cite{Brewer1979,mcpherson2001birds,Simsek2008}. 

\para{Research questions.} In this paper, we thus ask: How sensitive are different sampling techniques with respect to conserving the ranking of nodes and the visibility of groups in synthetic and empirical social networks with (i) different minority and majority group proportions, and (ii) various levels of homophily?

\begin{figure*}
    \centering
    \includegraphics[width=0.8\textwidth]{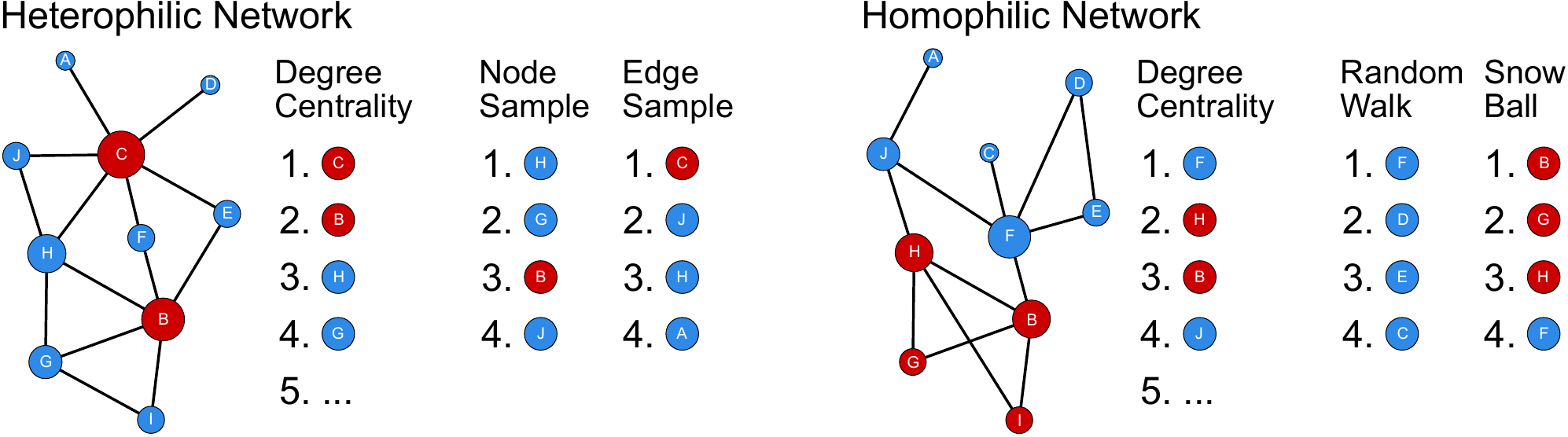}
    \caption{\emph{Illustration.} This example shows a heterophilic and a homophilic network with a red minority and a blue majority group. We illustrate that sampling methods may differ in their ability to preserve the \emph{visibility of the minority group} when ranking sampled nodes by their degree centrality.
    }
    \vspace{-1em}
    \label{fig:toy}
\end{figure*}

\para{Methods and materials.}
We evaluate different sampling techniques (node sampling, edge sampling, random walk sampling, and snowball sampling) with respect to reflecting the ranking of nodes and the visibility of groups in network samples (see Figure \ref{fig:toy}).
Instead of putting the focus on the whole population as in previous work, we specifically \emph{focus on sub-populations (or groups)}; we call the larger group \emph{majority} and the smaller group \emph{minority}. Our work is guided by the intuition that \emph{an ideal sample would allow to accurately preserve the original degree centrality ranking of nodes, and therefore preserve the relative importance between nodes and groups}. That means, an ideal sample would not systematically rank nodes of one group higher and nodes of the other group lower than expected. This would be considered a \emph{biased sample or sampling error}. 

\newpage
We construct synthetic social networks and vary the structural mechanisms guiding the growth of the network (i.e., homophily, preferential attachment, and group sizes), to study the extent to which they impact the accuracy of samples. We additionally showcase observed artifacts on empirical networks.
Based on the obtained insights, we provide indicators of why samples might have issues with capturing expected group characteristics.

\para{Contributions and Findings.}
(i) We propose a method to measure the robustness of samples from networks with two attributes. %
(ii) Using synthetic and empirical networks, we provide evidence that different network sampling techniques have issues with capturing the expected centrality of nodes and the visibility of minority / majority groups in social networks.
(iii) We discuss network characteristics that lead to observed discrepancies and quantify the impact of relative group size differences and homophily on sampling errors.

\section{Background and related work}

Network analysis has long been plagued by issues of measurement error, usually in the form of missing data. Understanding the robustness of basic network measures is extremely important in order to assess the validity of network research. 
Prior research explored the impact of missing data on various network measures, but mainly focused on small sociometric networks \cite{Costenbader2003,Galaskiewicz1991}, small bipartite collaboration graphs \cite{Kossinets2006}, and random networks \cite{Borgatti2006,Kossinets2006}.

Smith and Moody \cite{Smith2013} extended this line of research and  analyzed four classes of network measures on $12$ relatively small ($<1000$ nodes) empirical networks. They found that larger, more centralized networks, are in general more robust to missing nodes at random, especially for centrality and centralization measures. 
This is plausible since random node deletion in a centralized network (with skewed degree distribution) is less likely to remove hubs since few of them exist. 
In our work, we do not explore the effect of random node deletion, but, compare different sampling methods. Node sampling is the opposite of random node deletion, since the randomly selected nodes are included in the sample. %
Our results throughout this paper show that random node sampling from centralized networks (heterophilic networks with very popular minority) does not only fail to capture the centralization of the network well (since we miss the hubs), but also fails to accurately capture the relative importance of groups.

Wang et al \cite{wang2012measurement} presented the first work that explores the sensitivity of different network measures with respect to missing data in two large online social networks and one random graph. 
They defined six different types of measurement errors (missing nodes, spurious nodes, missing edges, spurious edges, falsely aggregated nodes, and falsely disaggregated nodes) and simulated their effect on the complete network. Using Spearman rank correlation, the authors compared the list of nodes that is ranked based on the network measure in the original network with the one that is computed on the sample. 
The work finds support for Borgatti's findings \cite{Borgatti2006}, highlighting that different centrality measures are similarly robust to measurement errors. Interestingly, results show that more local network measures like clustering are more prone to missing data than more global measures such as centralities. Thus, the authors revised the general claim from past research that the more ``global'' a measure, the less resistant it is to measurement error.

Lee et al. \cite{lee2006statistical} analyzed scale-free networks and three empirical networks suggesting that network properties such as betweenness centrality or clustering are sensitive to the choice of sampling method.
Lee and Pfeffer \cite{Lee2015} explored the quality of sampling by comparing the node-level network scores induced from the sample and the original network. 
They used edge-sampling and focused on degree and betweenness centrality for two empirical communication networks.
Their results show that larger samples lead to high sampling accuracy and that centralized graphs in which fewer nodes enjoy higher attention offer more accurate samples when edge sampling is used. 
Our work extends their work, since we compare various sampling techniques and introduce groups and homophily.

Furthermore, Leskovec and Faloutsos \cite{leskovec2006sampling} showed that network properties are sensitive to the choice of the sampling method. However, they assessed the quality of a sample by comparing the shape of the distribution of a network measure (e.g., degree) in the sample with the original one using the Kolmogorov Smirnov Distance. This evaluation criterion is very different from what has been used in previous work and what we use in this work, since it does not take the accuracy of the ranking of nodes into account.

Most prior work shows that network estimates become more inaccurate with lower sample coverage, but there is a wide variability of these effects across different measures, network topologies and sampling errors.
To our best knowledge, most previous work neglected the existence of heterogeneous attributes in networks and did not analyze the interplay between mechanisms that impact the topology of a social network and the accuracy of sampling techniques. A mentionable exception is the work by Li and Ye \cite{Li2011} who explored the ratio of intra- and inter- group links 
in samples drawn from a sample of the follow-network of Twitter users.
Our work extends their work by systematically exploring the effect of group sizes and homophily on the visibility of individual nodes. 
Our work focuses on undirected networks, but the work by Huisman \cite{Huisman2009} provides a comparison of sample bias in directed and undirected versions of the same network.

\section{Methods}
\label{sec:methods}

In this work, we are interested in studying the accuracy of samples drawn from networks with unequally sized groups and various levels of homophily.
We (i) describe used sampling techniques and (ii) explain how we assess the accuracy of a sample.

\subsection{Sampling techniques}

Our goal is to sample $K$ nodes from the overall set of $N$ nodes in a network. 
As pointed out in \cite{leskovec2006sampling},  we can split sampling algorithms into three groups: methods based on randomly selecting nodes, randomly selecting edges, and exploration techniques
simulating random walks or virus propagation to find
a representative sample of nodes.
We focus on one sampling technique from each group:

\para{Random node sampling.} This is the most basic sampling technique where a random subset of $K$ nodes is selected. The sampled network then contains these $K$ nodes and all links between them. Random node sampling is e.g., used when a sample of individuals is first selected and then their contact behavior is observed. Numerous surveys and data collections use this method, e.g., measuring contact pattern among  high school students using wearable sensors \cite{sociopattern_highschool}.

\para{Random edge sampling.} This strategy randomly samples edges from the network and filters the complete network by sampled edges. To be consistent with the other sampling strategies, we successively sample edges until $K$ nodes are selected. The sampled network then contains these $K$ nodes and sampled links, but not those links between selected nodes that have not been sampled. 
Random edge sampling is commonly used to construct a social graph by using information about contacts---e.g., phone calls are sampled and a graph of callers and receivers is constructed \cite{hidalgo2008dynamics}.

\para{Snowball sampling.} In snowball sampling, we randomly sample one starting node and add all its neighbors as well as the neighbors' neighbors to the set of sampled nodes---i.e., two step snowball sampling. We repeat this until we have gathered $K$ nodes for the sample. If a full iteration does not catch $K$ nodes, we repeat the process again with a new randomly selected starting node. The sampled network then contains these $K$ nodes and all the links connecting them.
Traditionally, snowball sampling is used when the population under study is not easily accessible (e.g., to study homeless people or illegal immigrants). Indeed, the promise of the snowball sampling is to access hard-to-reach population \cite{atkinson2001accessing}.

\para{Random walk (RW) sampling.} This strategy samples nodes by walking through the network. The walker starts at a random node in the network and chooses in each step one out-going link randomly and traverses it. All visited nodes are then added to the sample until $K$ nodes have been added. A teleport probability can be set for teleporting to another random node in the network instead of traversing a link in this iteration; we use $0.15$ throughout this work. The sampled network then contains these $K$ nodes and all links between them. This technique of sampling is usually used in online social networks such as Facebook or Twitter, in which retrieving information about the whole population is overwhelming and computationally costly, but we can access and navigate the original network.

\subsection{Evaluation measures}
The ubiquity of sampled network data makes the understanding of the robustness of network measures crucial. Here, we focus on the most basic and widely used centrality measure: \emph{degree centrality} \cite{Freeman1979}. The degree centrality of a node is defined as the fraction of nodes it is connected to. 

Previous work explored the robustness of centrality measures in samples of networks without taking heterogeneous attributes of nodes into account. Therefore, simple rank correlation (see e.g., \cite{Costenbader2003,wang2012measurement,Lee2015,Smith2013}) and overlap measures (see e.g., \cite{Borgatti2006}) have been used to assess how well a sample captures the ranking of nodes according to various network measures.
In this work, we are interested in assessing how well a sample captures, on average, the overall position of nodes in the original network for each group of nodes separately.
That means, we aim to reveal if the positions of nodes in both groups are  equally well captured in a way that the relative group and node importance are preserved.

If we would compute the overall rank correlation (or overlap) between the two lists and ignore the group memberships, then the ranking of majority nodes would contribute more to the correlation coefficient (or overlap). 
A naive group-specific measure would be to compute a separate rank correlation (or overlap) for each group. However, this measure would only allow us to assess how well the relative importance of nodes within each group in the original network is preserved in the sample, but the relation between nodes across groups would be neglected.
Therefore, simple rank correlation or overlap measures cannot be used to assess whether the relevance of nodes and groups is accurately captured in a sample.

In this work we define \emph{an ideal sample as a sample that allows to accurately reconstruct the original degree centrality ranking of nodes and therefore preserves the relative importance between nodes and groups}. That means, an ideal sample does not systematically rank nodes of one group higher and nodes of the other group lower than expected. 
To assess the \emph{accuracy of the relative importance of nodes and groups}, we propose the following two evaluation measures.
Both evaluation measures focus on the top $k$ or top $k$ percent of the data, since (i) users focus on the first few results in ranked lists %
and (ii) the distribution of degree centralities are usually heavy tail distributions. %
Therefore, the contribution of disorders in the long tail (unpopular nodes) would dominate disorders in the head (popular nodes) if we would not limit our analysis to the head \cite{Webber2010}.

\para{Top $k$ bias.}
To assess \emph{the accuracy of group visibility in a sample}, we compare the fraction of minority nodes in the top $k$ nodes of a sample with its fraction in the top $k$ nodes of the complete network. \begin{equation}
bias_{topk} = expected_{topk} - observed_{topk} \\
\label{ea:visibility_accuracy}
\end{equation}
$Observed_{topk}$ refers to the fraction of minority nodes that we observe in the top $k$ nodes of the sample, while $expected_{topk}$ refers to the fraction of minority nodes in the top $k$ nodes of the original network.
As sample size grows, the observed fraction in the sample approaches the expected fraction.

\begin{figure*}[t!]
	\centering
	\subfloat{\includegraphics[width=\textwidth]{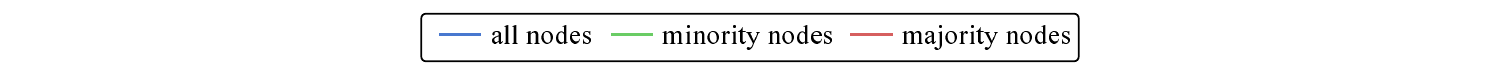}} \\ \vspace{-1em}
	\addtocounter{subfigure}{-1}
	\subfloat{\includegraphics[width=0.99\textwidth]{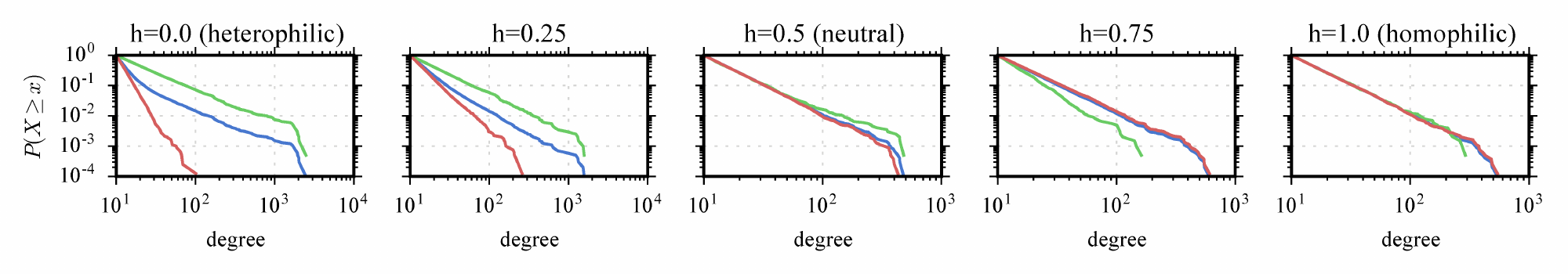}\label{<figure1>}}
	\vspace{-1em}
	\caption{\emph{Degree distribution of synthetic networks.} The average degree distribution of majority (80\% of nodes) and minority (20\% of nodes) in a synthetically generated preferential attachment network with various levels of homophily. One can see that the degree distributions are almost equal if homophily does not play a role ($h=0.5$). In heterophilic networks ($h<0.5$) the group-specific differences are much more pronounced than in homophilic networks ($h>0.5$).
	}
	 \vspace{-1em}
	\label{fig:homophily-degree}
\end{figure*}

\para{Normalized Cumulative Group Relevance (nCGR).}
The top $k$ ratio is a binary measure that does not take the importance of individual nodes into account. That means, we cannot measure how much lower the ranking of a node is in the sample compared to its ranking in the complete network.
To overcome this limitation, we first compute the relevance for each node $i$ by ranking nodes based on their centrality in the original network. The relevance of node $i$ is defined as the inverse rank that belongs to node $i$ normalized by the rank sum of all nodes ($N$) in the original network:
\begin{equation}
rel_{i} = \frac{inv\_rank_{i}}{\sum_{j=1}^{N}rank_{j}}
\end{equation}

The relevance shrinks linearly with the position of nodes in the list, but different weighting is possible. 
We compute for each group $g$ its cumulative group relevance ($CGR$) at rank k in the original ranked list and compare it with the cumulative relevance at rank k in the sample:
\begin{equation}
CGR_{topk} = \sum_{j=1}^{k}rel_{j \in g}
\end{equation}
\begin{equation}
nCGR_{topk} = \frac{CGR_{topk}(sample)+\epsilon}{CGR_{topk}(original)+\epsilon}
\end{equation}

The $nCGR_{topk}$ measures the extent to which the relevance of a group in the sample is above or below what we would expect from the original network with respect to the top $k$ nodes.
If e.g., this normalized cumulative group relevance for the minority is 2, then that means that the minority is twice as relevant in the sample than in the original network (for some top $k$). If it is 0.5 then the group is half as relevant in the sample than in the original network. If it is 1 then the group has equal relevance in the original network and the sample. 
We analyze the log of the normalized cumulative relevance since otherwise the measure is bound by zero; thus, the ideal nCGR is zero.
To avoid division by zero and logarithm of zero, we add a small $\epsilon=0.001$.

\section{Simulation Experiments}
\label{sec:synthetic}

We construct synthetic networks and explore the effect of homophily and group size on the accuracy of samples in a controlled environment. 
First, we describe the network model which we use to create synthetic network data and second, we discuss the accuracy of centrality measures in samples drawn from these networks using different sampling methods.

\begin{figure*}[t!]
	\centering
	\subfloat{\includegraphics[width=\textwidth]{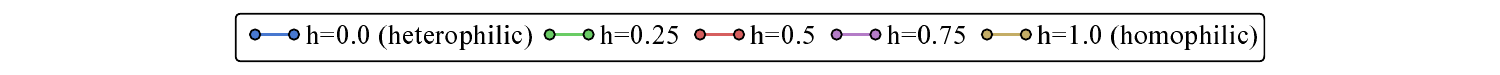}} \\ %
	\addtocounter{subfigure}{-1}
	\subfloat[Node sampling]{\includegraphics[width=0.25\textwidth]{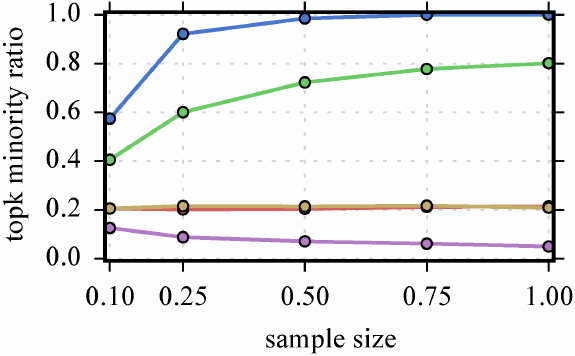}\label{<figure1>}}
	\subfloat[Edge sampling]{\includegraphics[width=0.25\textwidth]{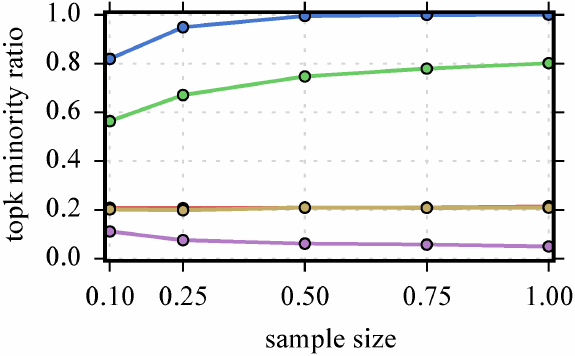}\label{<figure1>}}
	\subfloat[RW sampling]{\includegraphics[width=0.25\textwidth]{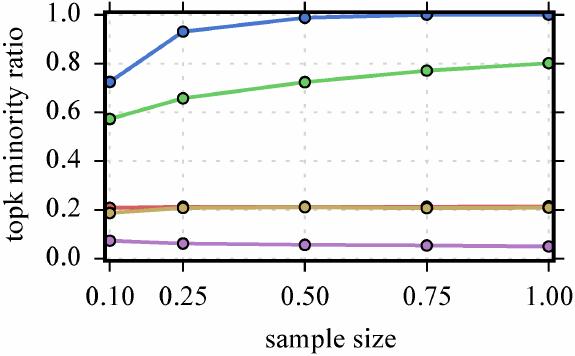}\label{<figure1>}} 
	\subfloat[Snowball sampling]{\includegraphics[width=0.25\textwidth]{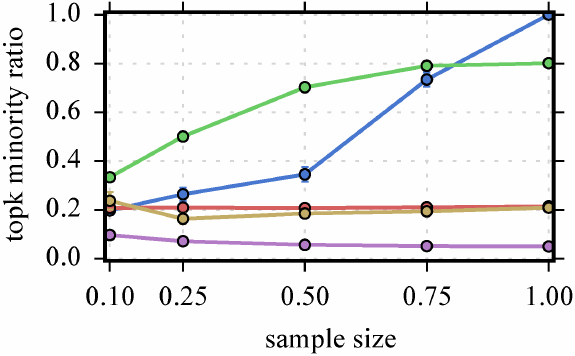}\label{<figure1>}}\\
	\vspace{-0.5em}
	\caption{\emph{Accuracy of group visibility in top-100.}  
		The y-axis visualizes the average percentage of minority nodes that show up in the top 100 nodes ranked by degree centrality computed on samples of different size.
		The x-axis depicts the respective sampling size and the last point 1.0 refers to the original network (see Eq.~ \ref{ea:visibility_accuracy}). 
		The lines refer to different homophily parameters that were used to generate the original network. Different subplots refer to different sampling techniques. 
		Overall, each point depicts the mean of 100 simulation runs based on 10 random network generation steps each having 10 sample steps; error bars mostly fall within the markers.
		One can see that in samples drawn from extreme and moderate heterophilic networks ($ 0.0 \leq h \leq 0.5$), the visibility of the minority is underestimated in small samples compared to what one would expect from the original network where sample size $=1.0$.
	}
	\vspace{-1em}
	\label{topk-ratio-synthetic}
\end{figure*}

\subsection{Synthetic network generators}

\noindent Preferential attachment (the tendency of nodes to connect to popular nodes) \cite{Yule1925,Barabasi99} and homophily (the tendency of nodes to connect to similar nodes) \cite{mcpherson2001birds,Simsek2008} have been extensively observed in many real-world social networks \cite{mislove2010you,fiore2005homophily,Crandall2008,Watts02identity} and information networks \cite{redner1998popular,Menczer2002}.
Homophily implies the existence of at least one fixed or mutable attribute (e.g., gender, ethnicity, education status).
Based on these attributes similarities between nodes can be defined.

We use an existing preferential attachment growth model with a homophily parameter that can be tuned and thus allows us to create networks with different levels of homophily and heterophily (see \cite{karimi_homophily,de2013scale} for details). 
The homophily parameter $h$ ranges between 0 to 1, $h \in [0,1]$, where $0$ means that nodes are only attracted by nodes that are dissimilar to them (heterophily), $1$ means nodes prefer to connect with similar nodes (homophily), and $0.5$ means that the link formation behavior is not driven by attributes.
All nodes of the same group share the same homophily parameter $h$, because they share the same attribute value and thus have the same distance to other groups with different attribute values.
We generate all synthetic networks with $10,000$ nodes and a fixed minority ratio of $20\%$ (except when noted otherwise).
An incoming node connects to $10$ nodes based on a specific homophily parameter and popularity (see \cite{karimi_homophily}).

Figure \ref{fig:homophily-degree} shows the degree distribution of both groups of nodes in networks that only vary in their degree of homophily. One can see that if we have two groups of unequal size and the network is heterophilic ($h<0.5$), the degree distributions of majority and minority differ the most. In fact, the fraction of high degree nodes that are part of the minority is much higher than it is for majority nodes. This is not surprising since the majority is attracted by the minority which therefore becomes an elite of powerful nodes in the network.  If the group membership does not play a role ($h=0.5$), the degree distributions of both groups are almost identical because only degree impacts the formation of edges and degree is equally distributed across groups. Also if the two groups are separated ($h=1.0$), the degree distributions are similar because both groups grow similarly and do not compete. The popularity of a group is bound by its size and therefore nodes with the highest degree are majority nodes.

If we compare the degree distribution of the two groups in a moderate heterophilic network with $h=0.25$ and a moderate homophilic network with $h=0.75$, we see that the differences between the degree distributions are more pronounced in the heterophilic case.
This asymmetric effect can be explained by the interplay between group size differences and homophily. 
The majority benefits from moderate homophily (e.g. $h=0.75$) more than from high homophily (e.g., $h=0.9$), because in high homophily conditions, their maximum degree is bound to the size of their group, while in moderate homophily conditions, sometimes also minority nodes will be attracted by the high degree of majority nodes.
Unlike the majority in the homophilic case, the minority in the heterophilic case benefits more from extreme heterophily (i.e., $h=0.0$) than from moderate heterophily (e.g. $h=0.25$). That is because in the extreme heterophily condition, all majority nodes are attracted by the minority, but in moderate heterophily condition sometimes the majority is attracted by high degree nodes which can also be part of their group.
So \emph{for the minority to gain popularity, it is better if they do not have to compete with the majority while the majority benefits from a competitive environment.}
In the next section, we will analyze how these group-specific differences in the degree distributions relate to sample biases.

\begin{figure*}[t!]
	\centering
	\subfloat{\includegraphics[width=\textwidth]{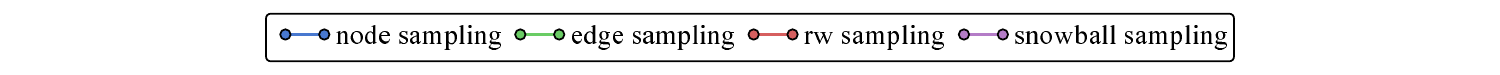}} \\ %
	\addtocounter{subfigure}{-1}
	\subfloat[Heterophilic ($h=0.25$)]{\includegraphics[width=0.25\textwidth]{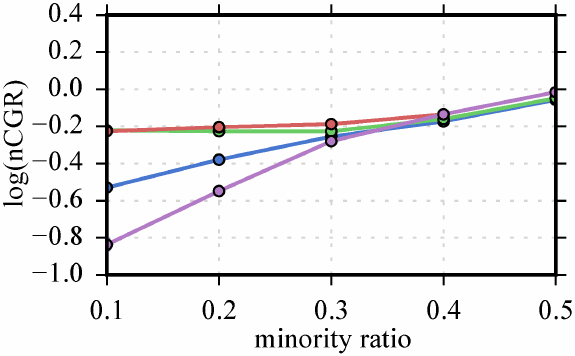}\label{<figure1>}}\hspace{1em}
	\subfloat[Neutral ($h=0.5$)]{\includegraphics[width=0.25\textwidth]{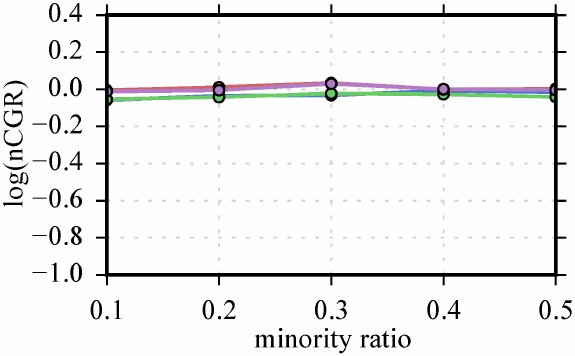}\label{<figure1>}} \hspace{1em}
	\subfloat[Homophilic ($h=0.75$)]{\includegraphics[width=0.25\textwidth]{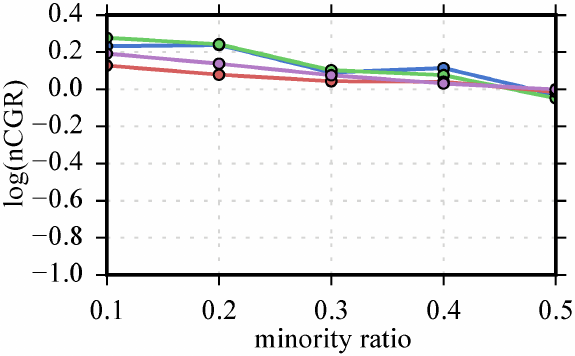}\label{<figure1>}}
	\vspace{-0.5em}
	\caption{\emph{Relative group size differences.} The y-axis shows the relevance of the minority group in the top 100 nodes of the sample network compared to the original network.
		The x-axis shows the relative size of the minority group. The sample size is 10\% of the original network.
		One can see that in samples drawn from heterophilic networks, the relevance of the minority is always underestimated; especially node and snowball sampling fail when group size differences are large in heterophilic networks.
		In homophilic networks the relevance of the minority is overestimated if the fraction of the minority group is very low.
		Node and edge sampling produce the most biased samples in this condition.
		Overall, we see that the more balanced the group sizes (0.5 means that 50\% of the nodes belong to minority) are, the more accurate the sample and the more similar the performance of different sampling techniques are.
		RW sampling performs best in all conditions and sampling errors are always higher in heterophilic networks than in homophilic ones.
	}
	\vspace{-1em}
	\label{fig:group-size-diff}
\end{figure*}

\begin{figure*}
	\centering
	\subfloat{\includegraphics[width=\textwidth]{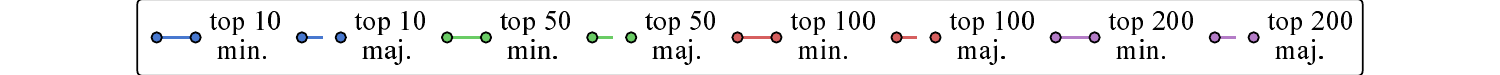}} \\ %
	\addtocounter{subfigure}{-1}
	\subfloat{\includegraphics[width=0.25\textwidth]{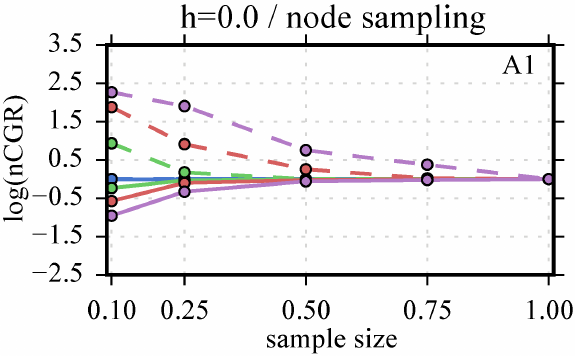}\label{<figure1>}}
	\subfloat{\includegraphics[width=0.25\textwidth]{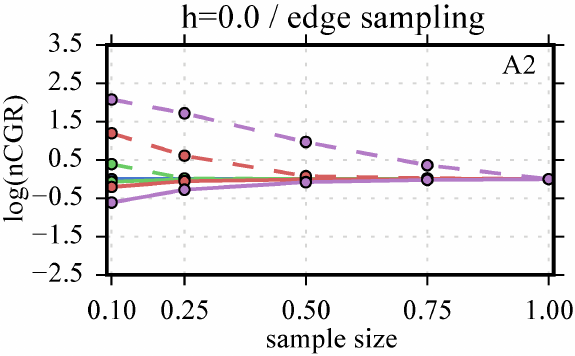}\label{<figure1>}}
	\subfloat{\includegraphics[width=0.25\textwidth]{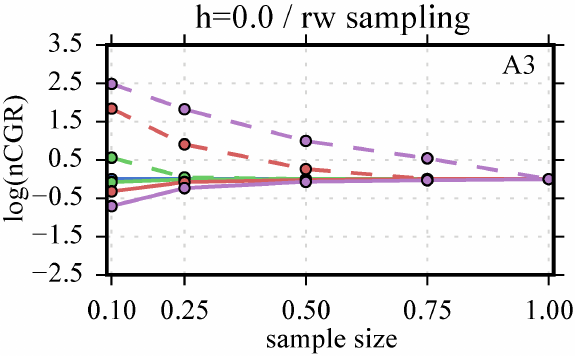}\label{<figure1>}}
	\subfloat{\includegraphics[width=0.25\textwidth]{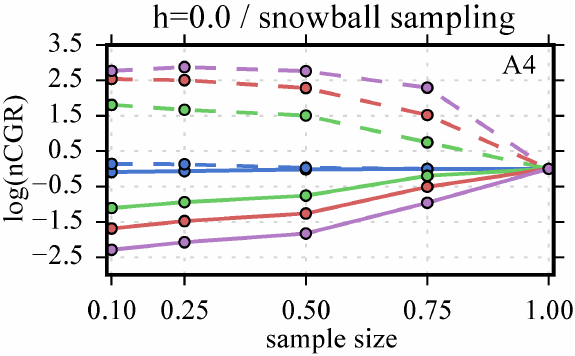}\label{<figure1>}} \\ %
	\subfloat{\includegraphics[width=0.25\textwidth]{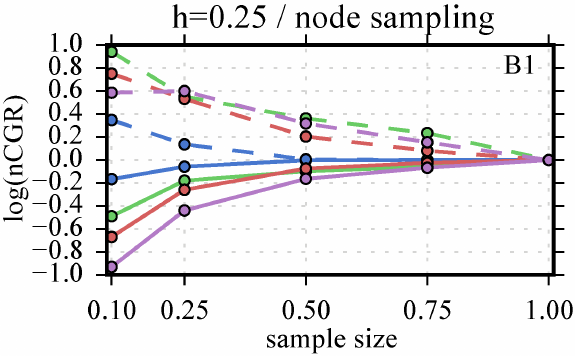}\label{<figure1>}}
	\subfloat{\includegraphics[width=0.25\textwidth]{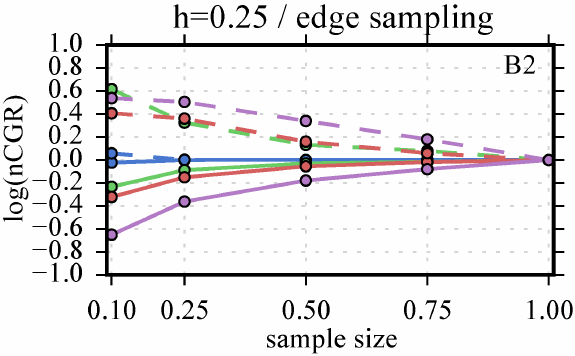}\label{<figure1>}}
	\subfloat{\includegraphics[width=0.25\textwidth]{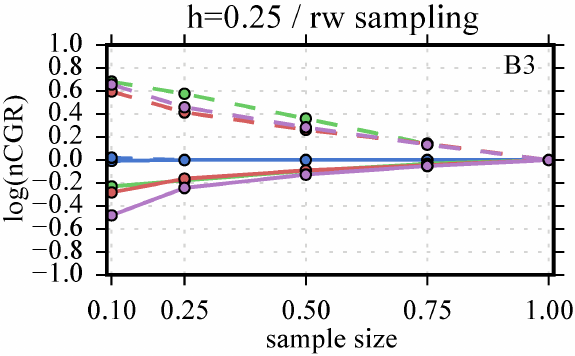}\label{<figure1>}}
	\subfloat{\includegraphics[width=0.25\textwidth]{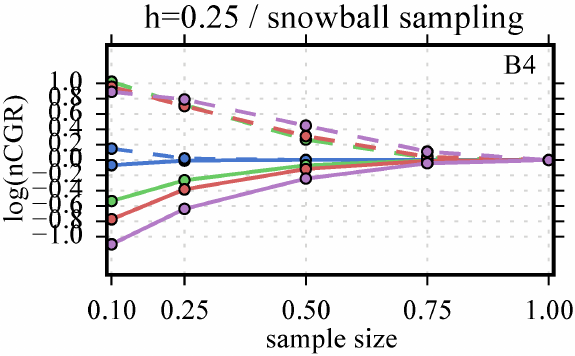}\label{<figure1>}} \\ %
	\subfloat{\includegraphics[width=0.25\textwidth]{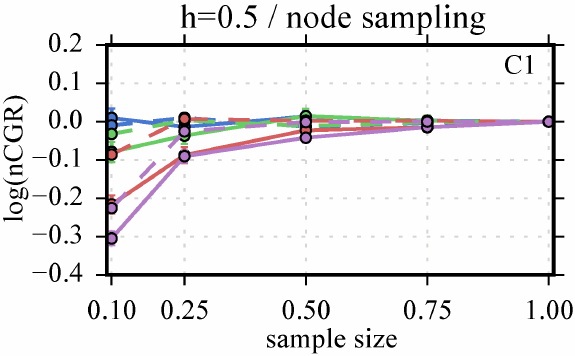}\label{<figure1>}}
	\subfloat{\includegraphics[width=0.25\textwidth]{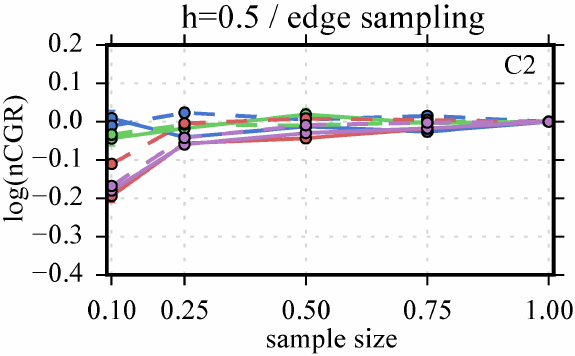}\label{<figure1>}}
	\subfloat{\includegraphics[width=0.25\textwidth]{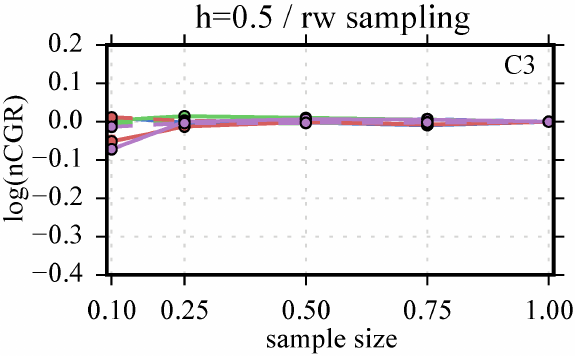}\label{<figure1>}}
	\subfloat{\includegraphics[width=0.25\textwidth]{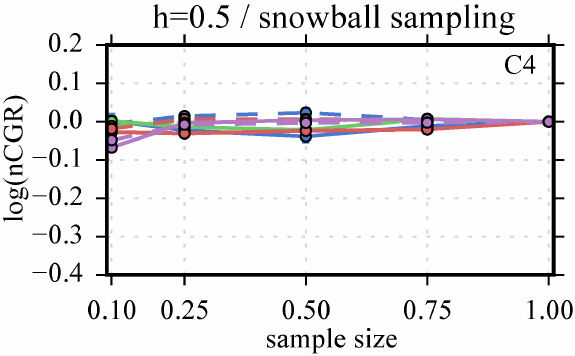}\label{<figure1>}} \\ %
	\subfloat{\includegraphics[width=0.25\textwidth]{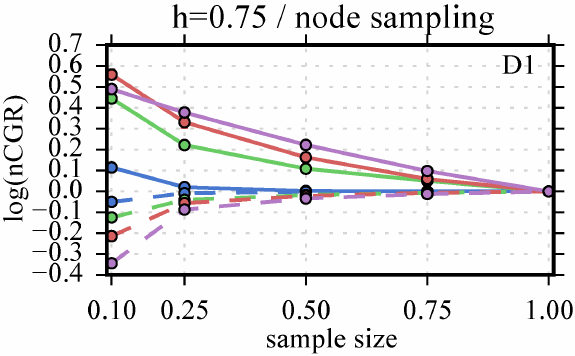}\label{<figure1>}}
	\subfloat{\includegraphics[width=0.25\textwidth]{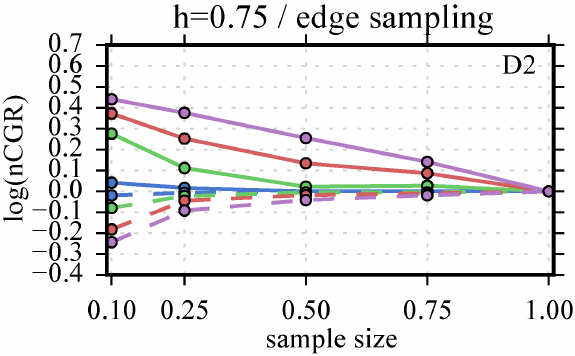}\label{<figure1>}}
	\subfloat{\includegraphics[width=0.25\textwidth]{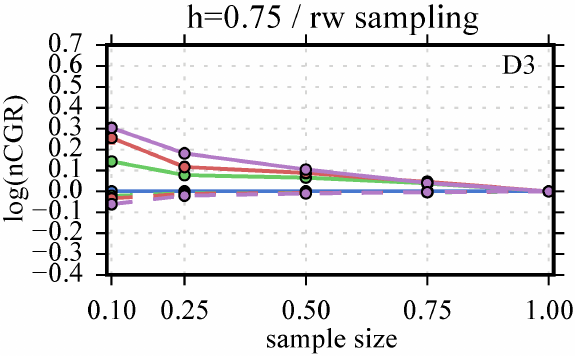}\label{<figure1>}}
	\subfloat{\includegraphics[width=0.25\textwidth]{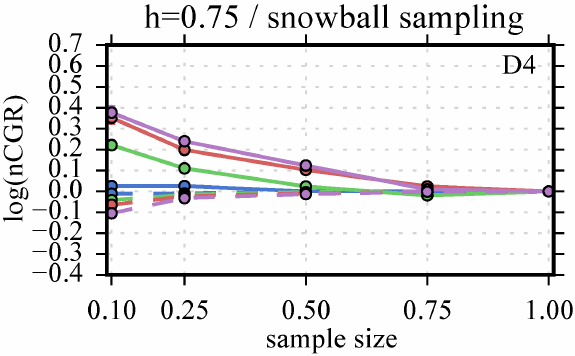}\label{<figure1>}} \\ %
	\subfloat{\includegraphics[width=0.25\textwidth]{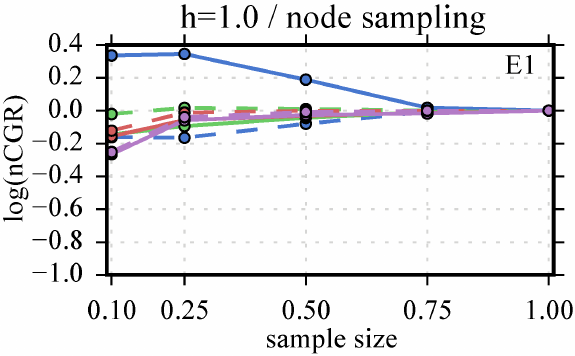}\label{<figure1>}}
	\subfloat{\includegraphics[width=0.25\textwidth]{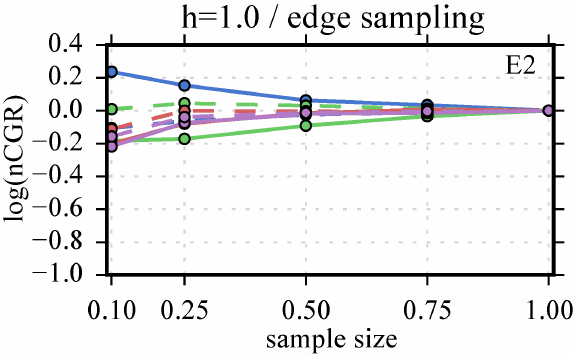}\label{<figure1>}}
	\subfloat{\includegraphics[width=0.25\textwidth]{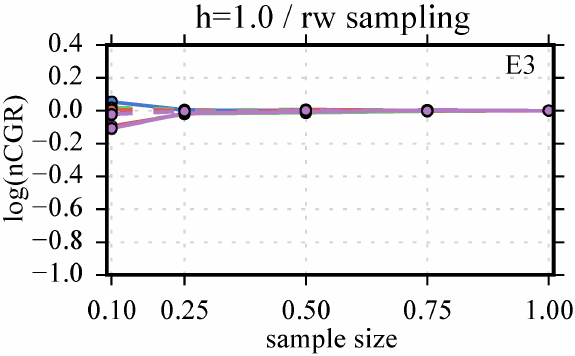}\label{<figure1>}}
	\subfloat{\includegraphics[width=0.25\textwidth]{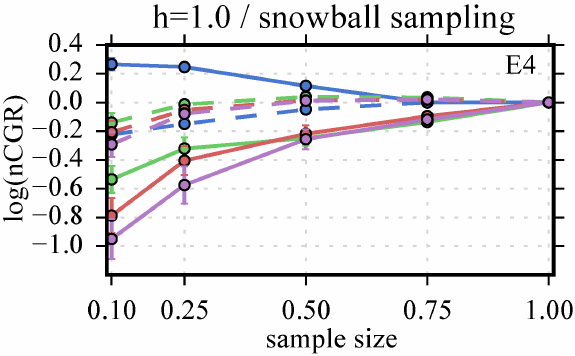}\label{<figure1>}} \\
	\vspace{-0.5em}
	\caption{\emph{Normalized Cumulative Group Relevance.} Each column depicts a different sampling technique, while each row refers to a different world for which the homophily level of the original network varies. The axis are aligned within each row, but not within each column, since the extent of error varies depending on the world.
	Again, each point refers to an average evaluation over $100$ total iterations.
	One can see that in extreme heterophilic networks (first row) the relevance of the majority is overestimated in small and also in larger sized samples, while the relevance of the minority is slightly underestimated especially in small sized samples. In extreme homophilic networks (last row), it is the other way around, however the extent to which the relevance of the minority is overestimated is smaller than the extent to which the relevance of the majority is overestimated in the extreme heterophilic case. Overall, random walk sampling produces the most accurate samples, followed by edge sampling. In samples based on node and snowball sampling, the relevance of the minority is usually underestimated, except in moderate homophilic networks (4th row).}
	\label{nCGR-synthetic}
\end{figure*}

\subsection{Sample bias in synthetic networks}
To assess sample bias, we generate synthetic networks, draw samples of varying size from them using different sampling techniques and assess the average visibility and relevance of different groups in samples. 
We repeat the random network generation process $10$ times and draw $10$ samples from each network; thus, in our evaluation, we report mean and standard error over $100$ samples.

Figure \ref{topk-ratio-synthetic} shows the visibility of the minority group in the top $100$ nodes in samples of different size which have been created via different sampling methods.
For example, in Figure \ref{topk-ratio-synthetic} (a), the point for the green line at an x-value of $0.10$ indicates that the top $100$ ranked nodes based on degree centrality in a $10\%$ sample from a moderate heterophilic network with $h=0.25$, contains on average around $40\%$ minority nodes.
We can compare this observed percentage with the expected percentage from the original network ($100\%$ sample). In this case, we would expect to see close to $80\%$ of minority nodes in the top $100$ nodes indicating that the minority is underrepresented in small samples drawn from moderate heterophilic networks with unbalanced group sizes using node sampling.

Results show that \emph{especially node and snowball sampling reduce the visibility of minority groups in the top $k$ list if samples are drawn from extreme and moderate heterophilic networks.} For node sampling, this is not surprising since all nodes have equal probability to be picked and therefore, a node's sampling probability is proportional to its group size. 
Snowball samples aggregate the 2-hop neighbourhood of randomly selected seed nodes which likely are majority nodes. 
Since most majority nodes are unpopular (skewed degree distribution), the probability for picking a majority node that has only a few minority nodes as neighbours is high.
Thus, we underestimate the visibility of the minority group in the top $k$. 
Figure \ref{fig:group-size-diff} shows that in the heterophilic network, the bias of node and snowball samples decreases linearly with decreasing group size difference. Note that group sizes are balanced if the minority ratio is $0.5$.
We further find that \emph{RW samples are very robust against relative size differences between groups in homophilic and heterophilic networks.}

In Figure \ref{nCGR-synthetic} we show to what extent the original relevance of each group is preserved in the sample. 
We find that \emph{in most cases the relevance of the minority is underestimated.} Only in moderate homophilic networks, minority is overrepresented. However, one needs to note that the extent with which the relevancy of the minority is overestimated in moderate homophilic networks ($h=0.75$, 4th row) is lower than the extent with which the relevancy of the majority is overestimated in moderate heterophilic networks ($h=0.25$, 2nd row).

Overall, we see that (i) the most accurate samples can be drawn from networks where homophily does not play a role, (ii) RW sampling performs best independent of the homophily conditions (see Figure \ref{topk-ratio-synthetic} and \ref{nCGR-synthetic}) and relative group size differences (see Figure \ref{fig:group-size-diff}), (iii) all sampling methods perform similar if group size differences are small, and (iv) the sampling error is always higher in heterophilic networks than in homophilic networks if the same sampling technique and group size differences are considered.

\begin{table*}[b!]
\centering
  \caption{ Coefficients of eight linear regression models, one for each sampling technique and sampling error measure. Each model was fitted to 3,200 observations (samples drawn from synthetically generated networks). The interaction term between group size difference and attribute influence is significant in node and snowball samples, but not in RW and edge samples. This indicates, the sampling error increases in node and snowball samples if the group size difference and the influence of attributes on the edge formation behavior are both increased. Edge and RW samples are rather robust against these factors. We compute the sampling error for lists of different length $k$ and control for the effect of $k$ in the model. The larger k, the higher the error. We also observe on average larger sampling errors on smaller sample sizes. \textit{Note:}  $^{**}p < 0.01$; $^{***}p < 0.001$.}
  \vspace{-0.5em}
  \small
  \begin{tabular}{|l|l|l|l|l|l|l|l|l|}
    \hline
     &
      \multicolumn{2}{c}{node sampling} &
      \multicolumn{2}{c}{snowball sampling} &
      \multicolumn{2}{c}{RW sampling} &
      \multicolumn{2}{c|}{edge sampling} \\
    & $nCGR_{topk}$ & $bias_{topk}$ & $nCGR_{topk}$ & $bias_{topk}$ & $nCGR_{topk}$ & $bias_{topk}$ & $nCGR_{topk}$ & $bias_{topk}$ \\
    \hline
    Intercept & 0.4064$^{***}$ & 0.0632 $^{**}$&  0.0810 & 0.0156 & 0.1057 & 0.0113 & 0.2117$^{***}$ & 0.0217\\
    \hline
    attr. infl. & 0.1203  & -0.1131  & 0.9866$^{**}$ & -0.0905 &  0.3491  &  0.0365  &  0.1153 &   -0.0506\\
    \hline
    grp. size diff & 0.0245 & -0.0602 & -0.4272 & -0.1106 & 0.0873& 0.0277& 0.0395 &  -0.0403 \\
    \hline
     attr. infl. : grp. size diff &  1.6851   &  0.5817$^{**}$ & 7.1846$^{***}$ & 1.7911$^{***}$& 1.1443 & 0.1396 & 1.0627 &   0.3464  \\
    \hline
   
     sample size & -0.8074$^{***}$ & -0.1233 $^{***}$& -1.1096$^{***}$ & -0.1278$^{***}$ & -0.5488$^{***}$& -0.0796$^{***}$ & -0.5131$^{***}$ & -0.0659$^{***}$  \\
    \hline
     top $k$ &  0.0014$^{***}$ &  0.0003$^{***}$&  0.0041$^{***}$ &  0.0006$^{***}$&  0.0017$^{***}$ &  0.0002$^{***}$& 0.0016$^{***}$ & 0.0003$^{***}$\\
    \hline \hline
    R2 & 0.281 & 0.135 & 0.418 & 0.428 & 0.225 & 0.152& 0.275 & 0.135 \\ \hline
  \end{tabular}
   \label{table:reg}
\end{table*}

\para{Regression analysis.}
To compare the impact of different factors on the sampling bias, we fit eight simple linear regression models, one model for each sampling technique and each error measure (top $k$ minority bias $bias_{topk}$ and the absolute sum of the normalized cumulative group relevance $nCGR_{topk}$ of the minority and the majority group). 
Each model was fitted to 3,200 observations (samples drawn from synthetically generated networks). 
Table \ref{table:reg} shows that across all sampling methods---perhaps not surprisingly---smaller samples lead to higher sampling errors and larger top $k$ lists lead to higher errors because the size of the network is constant. 
Interestingly, we see that only for node and snowball samples, the sampling error increases, if group size differences and the influence of the attribute on the edge formation behavior (i.e., the homophily parameter is closer to 0 or 1) increase. 
If only one of these factors changes, no significant effects on the sampling error can be observed, except for snowball samples. The bias of snowball samples also increases significantly if only homophily increases, because in extreme homophilic networks a snowball sample can only contain nodes of one group also if groups are of equal size. One can see that \emph{the sampling error of RW and edge samples cannot be explained by group size differences and homophily, which confirms our observation that these methods are rather robust against these factors.}

\begin{figure}[t!]
	\centering
\subfloat{\includegraphics[width=0.4\textwidth]{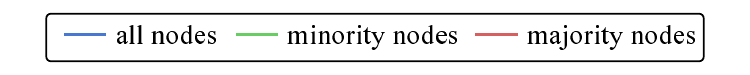}} \newline \vspace{-1.5em}
\subfloat{\includegraphics[width=0.49\textwidth]{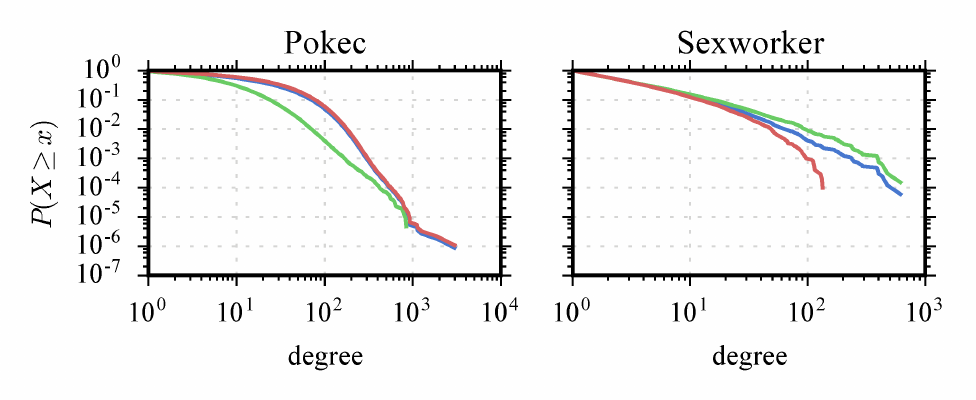}} 
\vspace{-1em}
	\caption{\emph{Degree distribution of empirical social networks (Pokec and Sexworker).} In the homophilic Pokec social network, nodes with the highest degree tend to belong to the majority (young users).  For the heterophilic sexworker network, the most popular nodes belong to the minority (women) since the majority (men) is attracted by the minority and the other way around. }
	\vspace{-1em}
	\label{fig:empirical-degree}
\end{figure}

\begin{figure*}[t!]
	\centering
    \subfloat{\includegraphics[width=\textwidth]{figlegend_dcg}} \\ %
	\addtocounter{subfigure}{-1}
	\subfloat[Pokec Node sample ]{\includegraphics[width=0.25\textwidth]{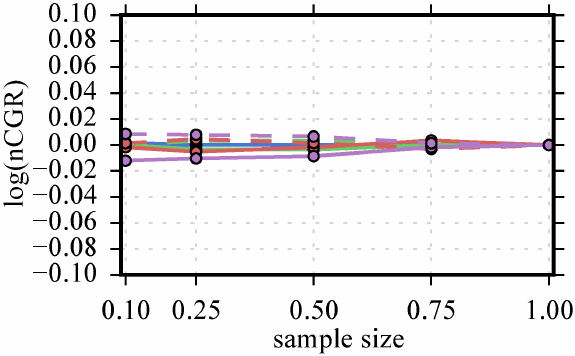}\label{ratio-nCGR-pokec-node}} 
	\subfloat[Pokec Edge sample]{\includegraphics[width=0.25\textwidth]{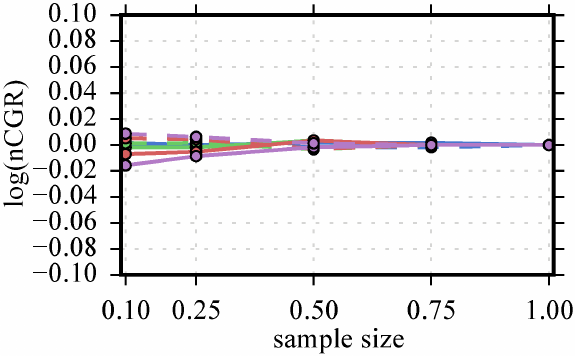}\label{ratio-nCGR-pokec-edge}} 
	\subfloat[Pokec RW sample ]{\includegraphics[width=0.25\textwidth]{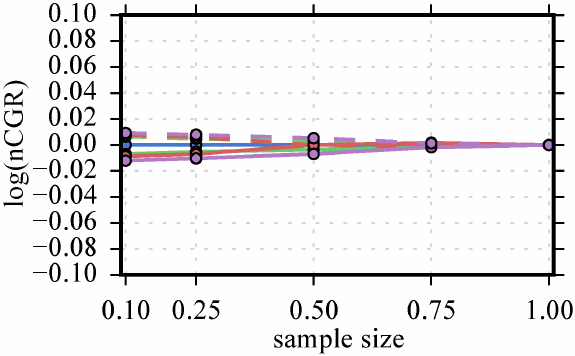}\label{ratio-nCGR-pokec-rw}} 
	\subfloat[Pokec Snowball sample]
	{\includegraphics[width=0.25\textwidth]{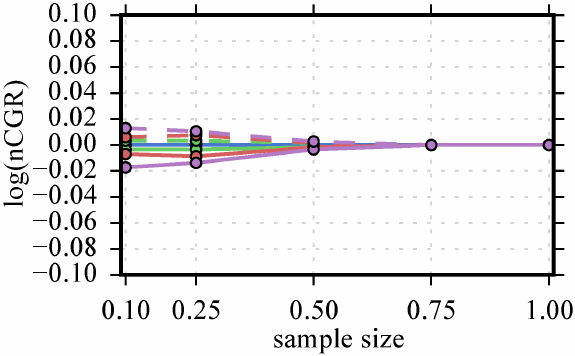}
	\label{ratio-nCGR-pokec-snow}} \\ 
	\addtocounter{subfigure}{-1}
	\subfloat[Sexworker Node sample]{\includegraphics[width=0.25\textwidth]{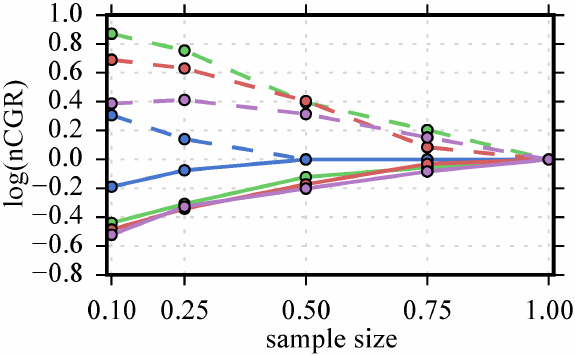}\label{ratio-nCGR-sexworker-node}}
	\subfloat[Sexworker Edge sample]{\includegraphics[width=0.25\textwidth]{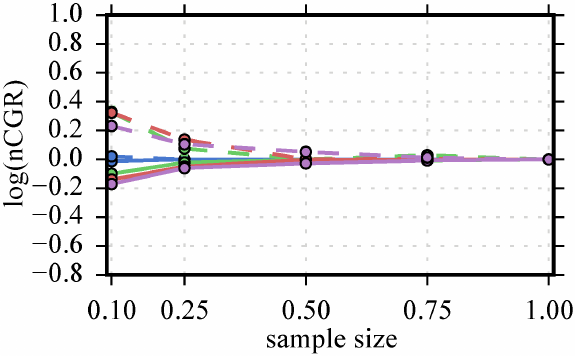}\label{ratio-nCGR-sexworker-edge}}
	\subfloat[Sexworker RW sample]{\includegraphics[width=0.25\textwidth]{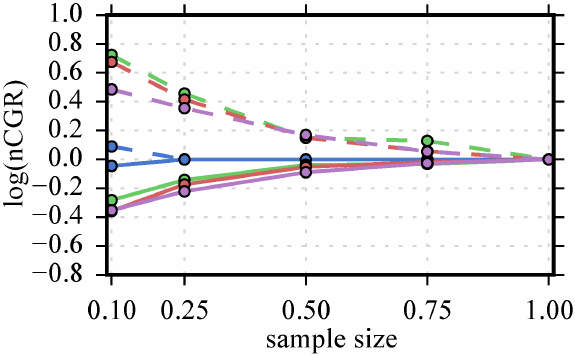}\label{ratio-nCGR-sexworker-rw}}
	\subfloat[Sexworker Snowball sample]{\includegraphics[width=0.25\textwidth]{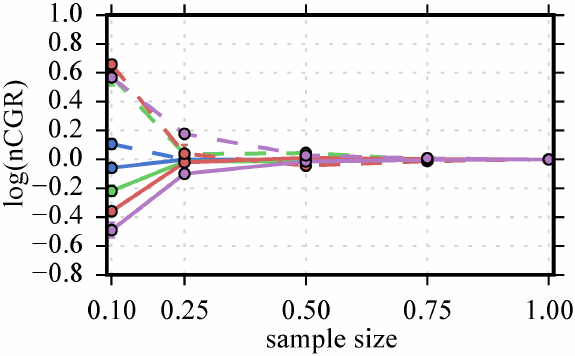}\label{ratio-nCGR-sexworker-snow}} \\ 
	\vspace{-0.5em}
\caption{\emph{Normalized Cumulative Group Relevance in empirical networks.} The first row refers to the Pokec network, the second to the Sexworker network. In samples drawn from the Pokec network, visibility and relevance of the minority correspond to what one would expect from the original network. Only for small top $k$, the minority is slightly more visible than expected if samples are generated via node, snowball or edge sampling. 
In samples drawn from the Sexworker network, we see that the minority (escorts) is very visible in the top 100 nodes ranked by degree centrality. Edge-based samples capture visibility of the minority in the original network best. The relevance of the majority is overestimated as also suggested by our model. Edge sampling produces the most accurate samples.
}
	\label{nCGR-empirical}
\end{figure*}

\section{Empirical Experiments}
\label{sec:empirical}

Next, we analyse two empirical networks and explore the accuracy of samples drawn from these networks. We describe the statistical properties of these networks and contrast empirical findings with the findings obtained from simulation.

\newpage
\subsection{Pokec social network}

\para{Dataset.}
We study publicly available data\footnote{\url{https://snap.stanford.edu/data/soc-pokec.html}} obtained from the most popular Slovakian social network ``Pokec'' \cite{takac2012data}. We added all friendship relations as undirected edges. The network contains $1,632,640$ nodes (users) and $22,301,602$ edges (friendship relations). The average degree of nodes is $27.32$, the global clustering coefficient is $0.0069$, and the graph diameter is $14$.

For our experiments, we focus on the age of actors in the social network. Eliminating all nodes without age information results in a network with $1,138,314$ nodes connected by $14,975,771$ edges. For coloring nodes as minority and majority, we take the $80\%$ percentile of the overall age distribution, and color all nodes with an age higher than this percentile as belonging to the \emph{minority} (old users), and all below as belonging to the \emph{majority}. This results in an age cut-off of $31$ years, meaning that the minority---$18.8\%$ of all nodes---captures the oldest users in the network. 
Overall, around $92\%$ of all edges in the network are between nodes of the same color---i.e., between two minority or two majority nodes. This exceeds the expectation of around $81.3\%$ if edges would form totally at random.
From that we can assert that the Pokec social network is moderately homophilic with respect to the defined age groups. 
Figure \ref{fig:empirical-degree} shows the degree distribution of young and old users. One can see that the most popular users are part of the majority.

\para{Results.}
Figure \ref{nCGR-empirical} shows that the visibility of the minority and the relevance of both groups is very well preserved in all samples. This is in line with what our model suggests for very homophilic networks (see Figure \ref{nCGR-empirical}).
Interestingly, random walk sampling produces the most accurate sample, which is also suggested by our model, especially for large relative groups size differences (see Figure \ref{fig:group-size-diff}(c)).

\subsection{Sexual contact network}

\para{Dataset.}
We use a network of claimed sexual contacts between Brazilian escorts (prostitutes) and sex buyers \cite{Rocha2011}.
The network consists of $16,730$ nodes 
(6,624 sex workers and 10,106 sex buyers)
and $50,632$ edges between them.
The minority of nodes with a share of around $40\%$ are sex workers, while the majority are sex buyers.
The network is fully bi-partite, meaning that sex workers only connect with sex buyers to capture sexual contacts. Consequently, all edges within the networks are between nodes of different color and thus, the network is 100\% heterophilic.
The degree distributions of minorities and majorities show that minorities are more popular than majorities (see Figure \ref{fig:empirical-degree}).
This is not surprising because the network is an example of an extreme heterophilic network since the majority nodes are attracted by the minority nodes and the other way around.

\para{Results.}
Figure \ref{nCGR-empirical} shows that the minority (escorts) are very visible in the top 100 nodes ranked by degree centrality also in samples of small size. Node-based samples are the most inaccurate samples, since they underestimate the visibility and relevance of the minority most.
Edge-based samples capture the visibility of the minority in the original network best if the original network is extremely heterophilic. 
Our model suggests that no large differences in the performance of different sampling techniques (as suggested by Figure \ref{fig:group-size-diff}) will exist because group size differences are rather small (40:60); but, edge and RW sampling will produce more accurate samples than node and snowball sampling.
Further, we can expect that all samples will underestimate the relevance of the minority. These expectations are confirmed empirically (cf. Figure \ref{nCGR-empirical}, bottom row).

\section{Discussion}
\label{sec:discussion}

If homophily (or heterophily) is the driving force behind the formation of edges in social networks with unbalanced attribute distributions, then the attribute and the degree of nodes become statistical dependent, i.e.,\linebreak $P(attribute|degree) \neq P(attribute) P(degree)$ and\linebreak $P(degree|attribute) \neq P(degree) P(attribute)$.  
Our work shows that if a statistical dependency between the network structure and the attribute of interest exists, all sampling methods introduce bias w.r.t. capturing the importance of nodes compared to when no relationship exists. %
However, not all sampling techniques are equally prone to group size differences and attribute influence on edge formation behavior which lead to statistical dependency between the network structure and the attribute of interest.
While sampling errors in node and snowball samples clearly increase if group size differences and attribute influence are increased, random walk and edge sampling are more robust against these factors.
This can be explained by the fact that e.g., random walk and edge sampling favor high degree nodes and aim to preserve the degree distribution of nodes. Therefore, systematic differences in the degree of nodes in different groups can, to some extent, be captured.
The sampling error in snowball samples also increases, if only the influence of attributes on the edge selection behavior increases (see Table \ref{table:reg}). This indicates, that even if group sizes are balanced, homophily or heterophily may cause problems in snowball samples.

Interestingly, the overestimation of the importance of a majority in heterophilic networks is more pronounced than the overestimation of the importance of minorities in homophilic networks. This can be explained by an asymmetry in the differences in degree distributions. In heterophilic networks, the difference between minority and majority degree distributions is larger than in a comparable homophilic network (same group sizes and similar impact of group membership on formation of edges).
Our observations from two real-world social networks confirm our simulation results and show that in heterophilic networks, the relevance of majority nodes is overestimated while in homophilic networks, it is slightly underestimated.

One limitation of our network generation model is that we limit it to two groups and that it assumes that all nodes in a group are equally active and behave equally homophilic or heterophilic. %
In real world social networks, more groups and group-specific and individual behavioral differences can be present.
Future research is necessary to study the effect of group-specific activity difference and asymmetric homophilic behavior and needs to explore the presence of multiple groups.
Furthermore, we focus on one specific network measure and undirected networks warranting further explorations about the accuracy of various network measures in samples drawn from directed networks.
Our work can be extended to more than one binary attribute by simply defining a similarity function that takes several attributes into account.

\section{Conclusions}
In summary, our work shows that the combination of two factors leads to sampling error in social networks with attributes: (i) group size differences and (ii) homophily.

If unequal sized groups are present, random walk sampling always leads to the most accurate samples---independent of the level of homophily.
The sampling error is always larger if samples are drawn from heterophilic networks with unequally sized groups compared to homophilic ones. 
In heterophilic networks with unbalanced groups, \emph{random walk} and \emph{edge sampling} perform similar well, while in homophilic networks \emph{edge sampling} produces more biased samples than \emph{random walk sampling}.
This can be explained by the fact that in homophilic networks edge sampling overestimates the importance of minority nodes, since minority nodes with high degree are more likely to be selected.
Edge samples only include sampled edges, but not all other edges between selected node. Therefore, the difference in degree between minority and majority nodes can be skewed.
Most sampling techniques produce accurate samples if the groups are of equal size. Only snowball samples can also be biased if homophily is a driving force behind the edge formation of nodes that belong to two equally sized groups.

Since researchers often do not have information about group size differences and homophily in the original network, random walk sampling is a robust choice. However, researchers cannot always choose their sampling method freely. Therefore, our results provide important guidance in estimating which groups will be over- or underestimated in samples drawn from social networks with unequally sized groups and various level of homophily. It is our hope that the research presented in this paper motivates more research into sampling from social networks with attributes.

\para{Acknowledgements.} We want to thank Robert West for valuable discussions and input to this work.
\clearpage
\begingroup
\raggedright
\sloppy
\balance
\bibliographystyle{abbrv}

\begin{thebibliography}{}

\end{thebibliography}


\begin{thebibliography}{10}

\bibitem{atkinson2001accessing}
R.~Atkinson and J.~Flint.
\newblock Accessing hidden and hard-to-reach populations: Snowball research
  strategies.
\newblock {\em Social research update}, 33(1):1--4, 2001.

\bibitem{Barabasi99}
A.-L. Barab{\'a}si and R.~Albert.
\newblock Emergence of scaling in random networks.
\newblock {\em Science}, 286(5439):509--512, 1999.

\bibitem{bearman2004chains}
P.~S. Bearman, J.~Moody, and K.~Stovel.
\newblock Chains of affection: The structure of adolescent romantic and sexual
  networks1.
\newblock {\em American journal of sociology}, 110(1):44--91, 2004.

\bibitem{Borgatti2006}
C.~K. Borgatti, S.P. and D.~Krackhardt.
\newblock Robustness of centrality measures under conditions of imperfect data.
\newblock {\em Social Networks}, 28(1):124–136, 2006.

\bibitem{Brewer1979}
M.~B. Brewer.
\newblock In-group bias in the minimal intergroup situation: {{A}}
  cognitive-motivational analysis.
\newblock {\em Psychological Bulletin}, 86(2):307--324, 1979.

\bibitem{Costenbader2003}
E.~Costenbader and T.~W. Valente.
\newblock {The stability of centrality measures when networks are sampled}.
\newblock {\em Social Networks}, 25(4):283--307, Oct. 2003.

\bibitem{Crandall2008}
D.~Crandall, D.~Cosley, D.~Huttenlocher, J.~Kleinberg, and S.~Suri.
\newblock Feedback effects between similarity and social influence in online
  communities.
\newblock In {\em Proceedings of the 14th ACM SIGKDD International Conference
  on Knowledge Discovery and Data Mining}, KDD '08, pages 160--168, New York,
  NY, USA, 2008. ACM.

\bibitem{de2013scale}
M.~L. de~Almeida, G.~A. Mendes, G.~M. Viswanathan, and L.~R. da~Silva.
\newblock Scale-free homophilic network.
\newblock {\em The European Physical Journal B}, 86(2):1--6, 2013.

\bibitem{fiore2005homophily}
A.~T. Fiore and J.~S. Donath.
\newblock Homophily in online dating: when do you like someone like yourself?
\newblock In {\em CHI'05 Extended Abstracts on Human Factors in Computing
  Systems}, pages 1371--1374. ACM, 2005.

\bibitem{Freeman1979}
L.~C. Freeman.
\newblock Centrality in social networks: Conceptual clarification.
\newblock {\em Social Networks}, 1(3):215--239, 1979.

\bibitem{Galaskiewicz1991}
J.~Galaskiewicz.
\newblock {Estimating point centrality using different network sampling
  techniques}.
\newblock {\em Social Networks}, 13(4):347--386, Dec. 1991.

\bibitem{hidalgo2008dynamics}
C.~A. Hidalgo and C.~Rodriguez-Sickert.
\newblock The dynamics of a mobile phone network.
\newblock {\em Physica A: Statistical Mechanics and its Applications},
  387(12):3017--3024, 2008.

\bibitem{Huisman2009}
M.~Huisman.
\newblock Imputation of missing network data: some simple procedures.
\newblock {\em Social Structure}, 10(1):1--29, 2009.

\bibitem{karimi_homophily}
F.~Karimi, M.~Génois, C.~Wagner, P.~Singer, and M.~Strohmaier.
\newblock Visibility of minorities in social networks.
\newblock {\em arXiv:1702.00150}, 2017.

\bibitem{Kossinets2006}
G.~Kossinets.
\newblock Effects of missing data in social networks.
\newblock {\em Social Networks}, 28:247--268, 2006.

\bibitem{Lee2015}
J.~Lee and J.~Pfeffer.
\newblock Estimating centrality statistics for complete and sampled networks:
  Some approaches and complications.
\newblock In {\em 48th Hawaii International Conference on System Sciences,
  {HICSS} 2015, Kauai, Hawaii, USA, January 5-8, 2015}, pages 1686--1695, 2015.

\bibitem{lee2006statistical}
S.~H. Lee, P.-J. Kim, and H.~Jeong.
\newblock Statistical properties of sampled networks.
\newblock {\em Physical Review E}, 73(1):016102, 2006.

\bibitem{leskovec2006sampling}
J.~Leskovec and C.~Faloutsos.
\newblock Sampling from large graphs.
\newblock In {\em Proceedings of the 12th ACM SIGKDD international conference
  on Knowledge discovery and data mining}, pages 631--636. ACM, 2006.

\bibitem{Li2011}
J.-Y. Li and M.-Y. Yeh.
\newblock On sampling type distribution from heterogeneous social networks.
\newblock In {\em Proceedings of the 15th Pacific-Asia Conference on Advances
  in Knowledge Discovery and Data Mining - Volume Part II}, PAKDD'11, pages
  111--122, Berlin, Heidelberg, 2011. Springer-Verlag.

\bibitem{sociopattern_highschool}
R.~Mastrandrea, J.~Fournet, and A.~Barrat.
\newblock Contact patterns in a high school: A comparison between data
  collected using wearable sensors, contact diaries and friendship surveys.
\newblock {\em PLoS ONE}, 10(9):e0136497, 09 2015.

\bibitem{mcpherson2001birds}
M.~McPherson, L.~Smith-Lovin, and J.~M. Cook.
\newblock Birds of a feather: Homophily in social networks.
\newblock {\em Annual Review of Sociology}, 27(1):415--444, 2001.

\bibitem{Menczer2002}
F.~Menczer.
\newblock Growing and navigating the small world web by local content.
\newblock {\em Proceedings of the National Academy of Sciences},
  99(22):14014--14019, 2002.

\bibitem{mislove2010you}
A.~Mislove, B.~Viswanath, K.~P. Gummadi, and P.~Druschel.
\newblock You are who you know: inferring user profiles in online social
  networks.
\newblock In {\em Proceedings of the third ACM international conference on Web
  search and data mining}, pages 251--260. ACM, 2010.

\bibitem{redner1998popular}
S.~Redner.
\newblock How popular is your paper? an empirical study of the citation
  distribution.
\newblock {\em European Physical Journal B}, 4(2):131--134, 1998.

\bibitem{Rocha2011}
L.~E.~C. Rocha, F.~Liljeros, and P.~Holme.
\newblock {Simulated Epidemics in an Empirical Spatiotemporal Network of 50,185
  Sexual Contacts}.
\newblock {\em PLoS Computational Biology}, 7(3), Mar. 2011.

\bibitem{shrum1988friendship}
W.~Shrum, N.~H. Cheek~Jr, and S.~MacD.
\newblock Friendship in school: Gender and racial homophily.
\newblock {\em Sociology of Education}, pages 227--239, 1988.

\bibitem{Simsek2008}
{\"O}.~{\c{S}}im{\c{s}}ek and D.~Jensen.
\newblock Navigating networks by using homophily and degree.
\newblock {\em Proceedings of the National Academy of Sciences},
  105(35):12758--12762, 2008.

\bibitem{Smith2013}
J.~A. Smith and J.~Moody.
\newblock Structural effects of network sampling coverage i: Nodes missing at
  random.
\newblock {\em Social Networks}, 35(4):652--668, 2013.

\bibitem{takac2012data}
L.~Takac and M.~Zabovsky.
\newblock Data analysis in public social networks.
\newblock In {\em International Scientific Conference and International
  Workshop Present Day Trends of Innovations}, pages 1--6, 2012.

\bibitem{wang2012measurement}
D.~J. Wang, X.~Shi, D.~A. McFarland, and J.~Leskovec.
\newblock Measurement error in network data: A re-classification.
\newblock {\em Social Networks}, 34(4):396--409, 2012.

\bibitem{Watts02identity}
D.~J. Watts, P.~S. Dodds, and M.~E.~J. Newman.
\newblock Identity and search in social networks.
\newblock {\em Science}, 296:1302--1305, 2002.

\bibitem{Webber2010}
W.~Webber, A.~Moffat, and J.~Zobel.
\newblock A similarity measure for indefinite rankings.
\newblock {\em ACM Transactions on Information Systems}, 28(4):1--38, Nov.
  2010.

\bibitem{Yule1925}
G.~U. Yule.
\newblock A mathematical theory of evolution, based on the conclusions of dr.
  j. c. willis, f.r.s.
\newblock {\em Philosophical Transactions of the Royal Society of London.
  Series B, Containing Papers of a Biological Character}, 213:pp. 21--87, 1925.

\end{thebibliography}

\endgroup

\end{document}